\documentclass{article}
\usepackage{spconf,amsmath,graphicx,hyperref}
\usepackage{makecell}
\usepackage{cite}
\usepackage{amsmath,amssymb,amsfonts}
\usepackage{float}
\usepackage{booktabs}
\usepackage{multirow}
\usepackage{enumitem}
\usepackage{algorithmic}
\usepackage{graphicx}
\usepackage{textcomp}
\usepackage{xcolor}
\usepackage{etoolbox}

\title{Lightweight Speech Enhancement Guided Target Speech Extraction in Noisy Multi-Speaker Scenarios}
\name{Ziling Huang$^{1,2 \dagger}$, Junnan Wu$^{2}$, Lichun Fan$^{2}$,  Zhenbo Luo$^{2}$, Jian Luan$^{2}$, Haixin Guan$^{3}$, Yanhua  Long$^{1,3}$\sthanks{Yanhua Long is the corresponding author. This work was sponsored by Natural Science Foundation of Shanghai (Grant No.25ZR1401277).\\$^{\dagger}$ Work done during internship at MiLM Plus, Xiaomi Inc.}}
\address{
  $^1$Shanghai Normal University, Shanghai, China\\
  $^2$MiLM Plus, Xiaomi Inc., Beijing, China;\\
  $^3$Unisound AI Technology Co., Ltd. Beijing, China
  }

\begin{document}
\ninept
\maketitle
\begin{abstract}

Target speech extraction (TSE) has achieved strong performance in relatively simple conditions 
such as one-speaker-plus-noise and two-speaker mixtures, but its performance remains unsatisfactory 
in noisy multi-speaker scenarios. To address this issue, we introduce a lightweight speech enhancement model, 
GTCRN, to better guide TSE in noisy environments. Building on our competitive previous 
speaker embedding/encoder-free framework SEF-PNet, we propose two extensions: LGTSE and D-LGTSE. 
LGTSE incorporates noise-agnostic enrollment guidance by denoising the input noisy speech before 
context interaction with enrollment speech, thereby reducing noise interference. 
D-LGTSE further improves system robustness against speech distortion by leveraging denoised speech 
as an additional noisy input during training, expanding the dynamic range of noisy conditions and 
enabling the model to directly learn from distorted signals. Furthermore, we propose a two-stage training strategy, 
first with GTCRN enhancement-guided pre-training and then joint fine-tuning, to fully exploit model potential.
Experiments on the Libri2Mix dataset demonstrate significant improvements of 
0.89 dB in SISDR, 0.16 in PESQ, and 1.97\% in STOI, validating the effectiveness of our approach. 
Our code is publicly available at \url{https://github.com/isHuangZiling/D-LGTSE}.

\end{abstract}


\begin{figure*}[ht]
\centering
\includegraphics[width=0.9\textwidth]{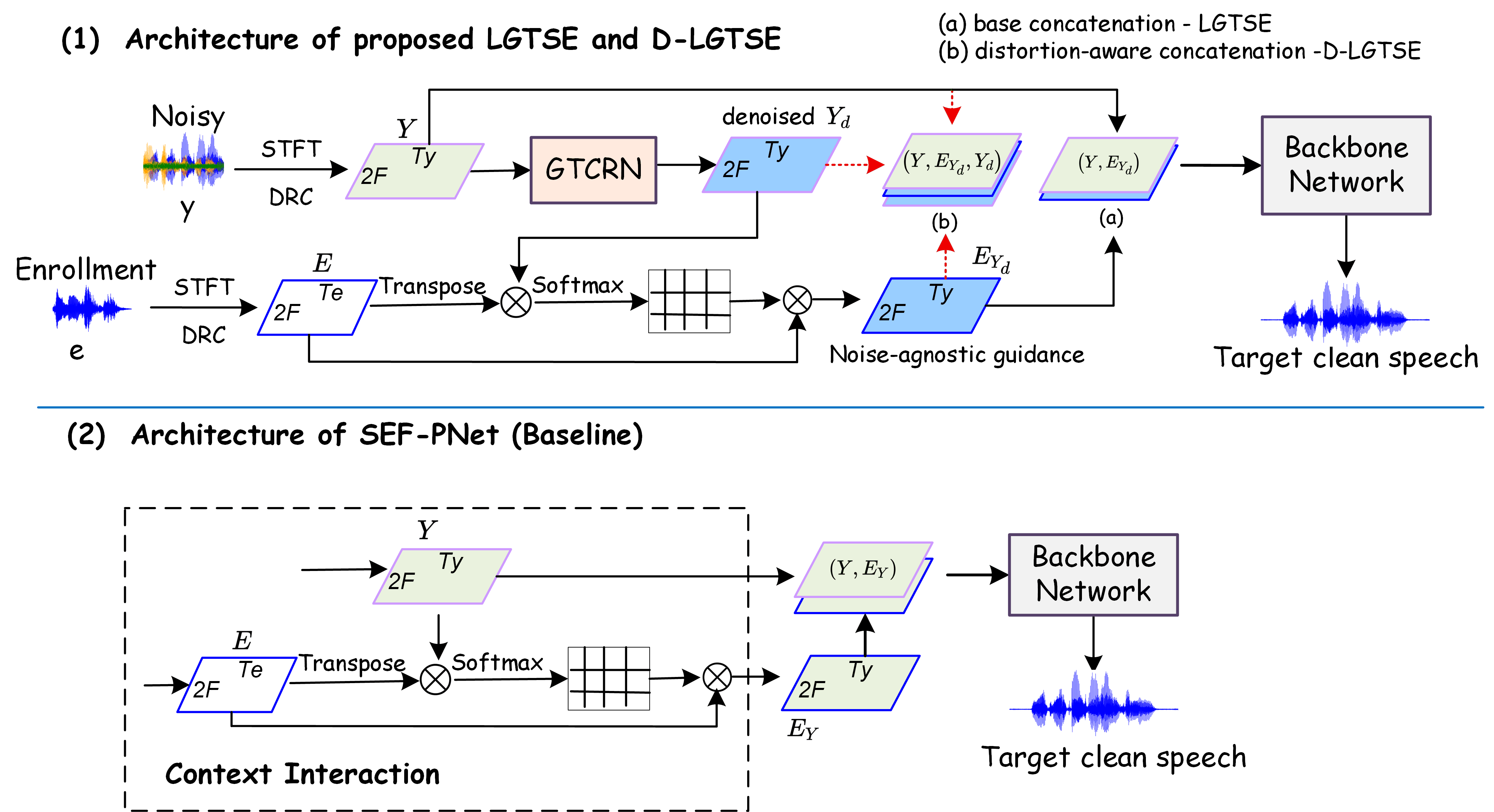}
\vspace{-0.3cm}
\caption{Architecture of the proposed LGTSE and D-LGTSE, and the simplified SEF-PNet (baseline).}
\label{fig:fig1}
\vspace{-0.3cm}
\end{figure*}

\section{Introduction}
\label{sec:intro}

Target speech extraction (TSE) aims to extract the speech of a desired speaker from mixtures of 
interferers and background noise using an enrollment utterance, with applications in ASR, 
hearing aids, and speech communication. While recent methods perform well in simple 
conditions (e.g., one-speaker-plus-noise, two-speaker mixtures), performance degrades 
significantly in noisy multi-speaker scenarios. A key challenge lies in the quality of 
enrollment guidance, as noise-corrupted enrollment can severely mislead the model in 
identifying the target speaker.

Recent studies on enrollment-guided TSE can be grouped into three categories:
1) speaker embedding/encoder-based approaches \cite{pbsrnn,xsepformer,sdpccn,spex,spex+,td-speakerbeam,you2025investigation}, 
2) embedding/encoder-free methods \cite{sefnet,sefpnet,yangxue,hu2024smma,parnamaa2024personalized}, 
and 3) hybrid techniques that combine the two \cite{zhang2025multi}.
Embedding/encoder-based methods obtain target speaker embeddings 
via pretrained embedding models \cite{ecapa,resnet}, or jointly with the separation backbone 
using a self-designed speaker encoder \cite{sdpccn,mc-spex,tea-pse, tea-pse2.0, tea-pse3.0}, or using both 
simultaneously\cite{he2024hierarchical}. In contrast, embedding/encoder-free methods 
avoid explicit embeddings by directly modeling enrollment–mixture interactions, e.g., via iterative attention 
\cite{sefnet}, RNN state summarization \cite{veven}, or STFT-domain attention \cite{yangxue}. 
Hybrid methods \cite{zhang2025multi} combine both embeddings 
and direct interaction to provide richer guidance. 
While embedding-based and hybrid approaches have achieved impressive results in various TSE tasks, 
their large size and slow inference limit practical deployment.
Therefore, recent research has increasingly shifted toward embedding/encoder-free paradigms, 
such as CIE-mDPTNet \cite{yangxue}, SEF-Net \cite{sefnet} and SEF-PNet\cite{sefpnet}, 
which already demonstrate state-of-the-art performance.


Although recent TSE methods have achieved remarkable progress, they still perform poorly in noisy multi-speaker scenarios.
For embedding-free approaches, the simultaneous presence of noise and interferers greatly increases 
task difficulty. Noise not only hinders separation but also severely corrupts enrollment guidance, 
often leading to target speech distortion. To address this, \cite{yang2024target} proposed a 
jointly trained enhancer to reduce enrollment–noise similarity, however,
the resulting guidance remains noise-contaminated since enrollment speech still interacts with mixtures. 
Other works \cite{tea-pse,zhang2024two,liu24n_interspeech,deepfilternet,he20243s}, employ multi-stage coarse-to-fine extraction, 
improving robustness but at the cost of nearly doubling the parameters. These challenges highlight the urgent 
need for lightweight, noise-robust, and distortion-resistant 
TSE solutions that can operate effectively in real-world noisy multi-speaker environments.

In this paper, we build upon our previous competitive embedding-free backbone, SEF-PNet \cite{sefpnet}, 
and propose two enhanced frameworks for noisy multi-speaker conditions by integrating a lightweight denoiser, 
GTCRN \cite{rong2024gtcrn}. The denoiser not only pre-processes noisy speech to provide 
noise-agnostic enrollment guidance, but also generates distorted variants of speech that 
are leveraged for training data augmentation. Our main contributions are as follows:
\begin{itemize}[leftmargin=*, itemsep=1pt, parsep=1pt, topsep=1pt]

\item \textbf{LGTSE:} introduces a lightweight guidance scheme where context 
interaction is performed between GTCRN-denoised noisy speech and clean enrollment 
speech in an end-to-end manner. This avoids direct interaction with noisy mixtures 
and substantially improves TSE performance under noisy multi-speaker scenarios.

\item \textbf{D-LGTSE:} further extends LGTSE by exploiting the denoised outputs 
as additional distorted training data. This augmentation strategy enriches the acoustic 
variability of noisy inputs and exposes the model to distortion, thereby enhancing robustness.

\item \textbf{Two-stage Training:} adopts a progressive training strategy 
in which the denoiser is first pre-trained, followed by training the TSE backbone, 
and finally fine-tuning the entire system jointly. This staged optimization 
yields consistent performance improvements.

\end{itemize}

\section{Proposed Methods}
\label{sec:methods}

\subsection{Architecture}

As illustrated in Fig.~\ref{fig:fig1}, the proposed LGTSE and D-LGTSE architectures 
are shown in the upper part, while the simplified SEF-PNet baseline is demonstrated below.  
In both frameworks, the noisy speech and enrollment speech are fed into the backbone’s front-end 
\textit{context interaction} module to generate a guidance feature, which is concatenated with the 
noisy/denoised feature and passed into the backbone network to extract the target clean speech.

Compared to our previously proposed simplified SEF-PNet, 
LGTSE introduces a lightweight speech enhancement model (GTCRN) to 
denoise the noisy speech before interaction with the enrollment speech, effectively preventing noise 
contamination in the guidance. In addition, D-LGTSE further leverages the denoised speech as an extra 
distorted input during training, thereby expanding the dynamic range of noisy conditions and enabling 
distortion-aware training. This operation further improves the model's robustness against speech 
distortion. Details of key components in LGTSE and D-LGTSE are presented as follows.

\vspace{-0.2cm}  
\subsection{Noise-agnostic Enrollment Guidance Extraction}

Directly performing context interaction between enrollment and noisy speech often 
leads to noise contamination in the enrollment-guided representation. To address this, 
both LGTSE and D-LGTSE introduce a noise-agnostic enrollment guidance extraction process, 
which produces a more robust target speaker representation and enables 
the backbone network to extract higher-quality target speech.

As shown in Fig.~\ref{fig:fig1}-(1), the inputs are enrollment speech and noisy speech, 
which are first transformed into complex time-frequency representations via short-time Fourier transform. 
Specifically, \(\mathbf{E} \in \mathbb{R}^{2F \times T_e}\) and \(\mathbf{Y} \in \mathbb{R}^{2F \times T_y}\), 
where \(2F\) denotes the concatenated real and imaginary parts along the frequency axis, and \(T_e\), \(T_y\) 
represent the number of frames of enrollment and noisy speech, respectively. Dynamic range compression \cite{drc} 
with compression factor \(\beta=0.5\) is applied on the magnitude spectrum:
\begin{equation}
\mathbf{E} = |\mathbf{E}|^\beta e^{j\theta_E}, \quad \mathbf{Y} = |\mathbf{Y}|^\beta e^{j\theta_Y}
\vspace{-0.1cm}
\end{equation}

If we follow the SEF-PNet baseline (Fig.~\ref{fig:fig1}-(2)), the enrollment representation $\mathbf{E}_Y$ is then obtained by directly performing context interaction between the enrollment and noisy features as:
\vspace{-0.1cm}
\begin{equation}
\mathbf{E}_Y = \mathbf{E} \times \mathrm{softmax}\left(\mathbf{E}^\mathrm{T} \times \mathbf{Y}\right)
\vspace{-0.1cm}
\end{equation}
where the softmax operation is applied along the enrollment time-frame dimension to measure the correlation 
between enrollment and noisy frames, the noise contained in \(\mathbf{Y}\) will inevitably contaminate the 
resulting guidance \(\mathbf{E}_Y\).

To mitigate this, our method first denoises the noisy speech with a lightweight speech enhancement model GTCRN, and then replaces \(\mathbf{Y}\) with the denoised \(\mathbf{Y}_d\) for the context interaction:
\vspace{-0.1cm}
\begin{equation}
\begin{aligned}
\mathbf{Y}_d &= \mathrm{GTCRN}(\mathbf{Y}), \\
\mathbf{E}_{Y_d} &= \mathbf{E} \times \mathrm{softmax}\left(\mathbf{E}^\mathrm{T} \times \mathbf{Y}_d\right)
\end{aligned}
\vspace{-0.1cm}
\end{equation}
By using the denoised feature in the context interaction, the target speaker guidance becomes noise-agnostic, effectively suppressing noise interference.

\vspace{-0.2cm}  
\subsection{Distortion-aware LGTSE (D-LGTSE)}

In multi-speaker TSE, extracted target speech often contains distortions. 
To improve robustness, D-LGTSE leverages the denoised spectrum $\mathbf{Y}_d$, 
which is not perfectly clean but mildly distorted, as an additional input during 
training. This exposes the model to distorted speech and increases acoustic variability. 
Three distortion-aware data usages are investigated as follows.

\textbf{Distortion-aware concatenation:}  As shown in Fig.~\ref{fig:fig1}-(1)-(b), 
unlike the \textit{base concatenation} used in LGTSE with only $\mathbf{Y}$ and $\mathbf{E}_{Y_d}$,
D-LGTSE concatenates the original noisy spectrum $\mathbf{Y}$, the denoised spectrum $\mathbf{Y}_d$, 
and the noise-agnostic guidance $\mathbf{E}_{Y_d}$ along the channel dimension. 
This fused representation is then fed into the backbone, enabling joint 
processing within a single forward pass.

\textbf{On-the-fly:} Each mini-batch $\mathcal{B}$ is enlarged by including both 
the original noisy and the denoised spectrums generated on-the-fly:
\vspace{-0.1cm}
\begin{equation}
\mathcal{B} = \{(\mathbf{Y}_i,  \mathbf{E}_{Y_{d}}^{i}),\mathbf{Y}^{i}_{\text{target}}\}_{i=1}^N \cup \{(\mathbf{Y}_{d}^{i}, \mathbf{E}_{Y_{d}}^{i}),\mathbf{Y}_{\text{target}}^{i}\}_{i=1}^N
\vspace{-0.1cm}
\end{equation}
where $N$ is the original mini-batch size,  $\mathbf{Y}_{\text{target}}^{i}$ 
is the clean target speech (ground-truth) corresponding to the $i$-th noisy sample. 
This lets the model process noisy and mildly distorted speech in parallel.

\textbf{Offline:} The entire noisy dataset $\mathcal{D}$ is first processed to 
obtain denoised dataset $\mathcal{D}_d$. The two datasets are then merged and 
shuffled to form a distortion-aware noisy training set $\mathcal{D}_\text{mix}$, 
which is paired with enrollment guidance and used to train the model following the 
LGTSE scheme.
\vspace{-0.1cm}
\begin{equation}
\mathcal{D}_\text{mix} = \mathrm{shuffle}(\mathcal{D} \cup \mathcal{D}_d)
\vspace{-0.1cm}
\end{equation}
The shuffle operation encourages the model to generalize better by exposing it to 
diverse noisy–denoised pairings. Moreover, compared with distortion-aware concatenation, 
this offline strategy reduces both computation cost and inference latency.

\vspace{-0.2cm}  
\subsection{Two-stage Training Strategy}

Both LGTSE and D-LGTSE adopt a two-stage training strategy to fully leverage the pretrained modules. In the first stage (pre-training), GTCRN is trained for speech enhancement only on noisy mixtures (e.g., `2-speaker + noise'), and its denoised outputs, together with enrollment speech, are used to pretrain the backbone network from scratch for TSE with noise-agnostic enrollment guidance. In the second stage (joint fine-tuning), GTCRN is unfrozen, and the entire system is fine-tuned jointly in an end-to-end manner, enabling the backbone to exploit denoised speech more effectively and thereby improving robustness and overall TSE performance.

The training objective minimizes the negative scale-invariant signal-to-distortion ratio (SI-SDR). For GTCRN, the ground-truth is the clean 2-speaker mixture speech $\mathbf{y}_{\text{clean}}$, and for the backbone, it is the target speech  $\mathbf{y}_{\text{target}}$. During end-to-end joint fine-tuning, both losses are combined:
\vspace{-0.1cm}
\begin{equation}
\mathcal{L} = -\text{SI-SDR}(\mathbf{y}_d, \mathbf{y}_{\text{clean}}) - \text{SI-SDR}(\mathbf{\hat{y}}_{\text{target}}, \mathbf{y}_{\text{target}})
\vspace{-0.1cm}
\end{equation}
where $\mathbf{y}_d$ is the GTCRN denoised speech, and $\mathbf{\hat{y}}_{\text{target}}$ is the backbone's target estimate. This jointly optimizes denoising and target speech extraction thus enhances the noisy multi-speaker TSE.

\begin{figure*}[t]
\centering
\setlength{\abovecaptionskip}{0cm}
\includegraphics[width=0.9\textwidth]{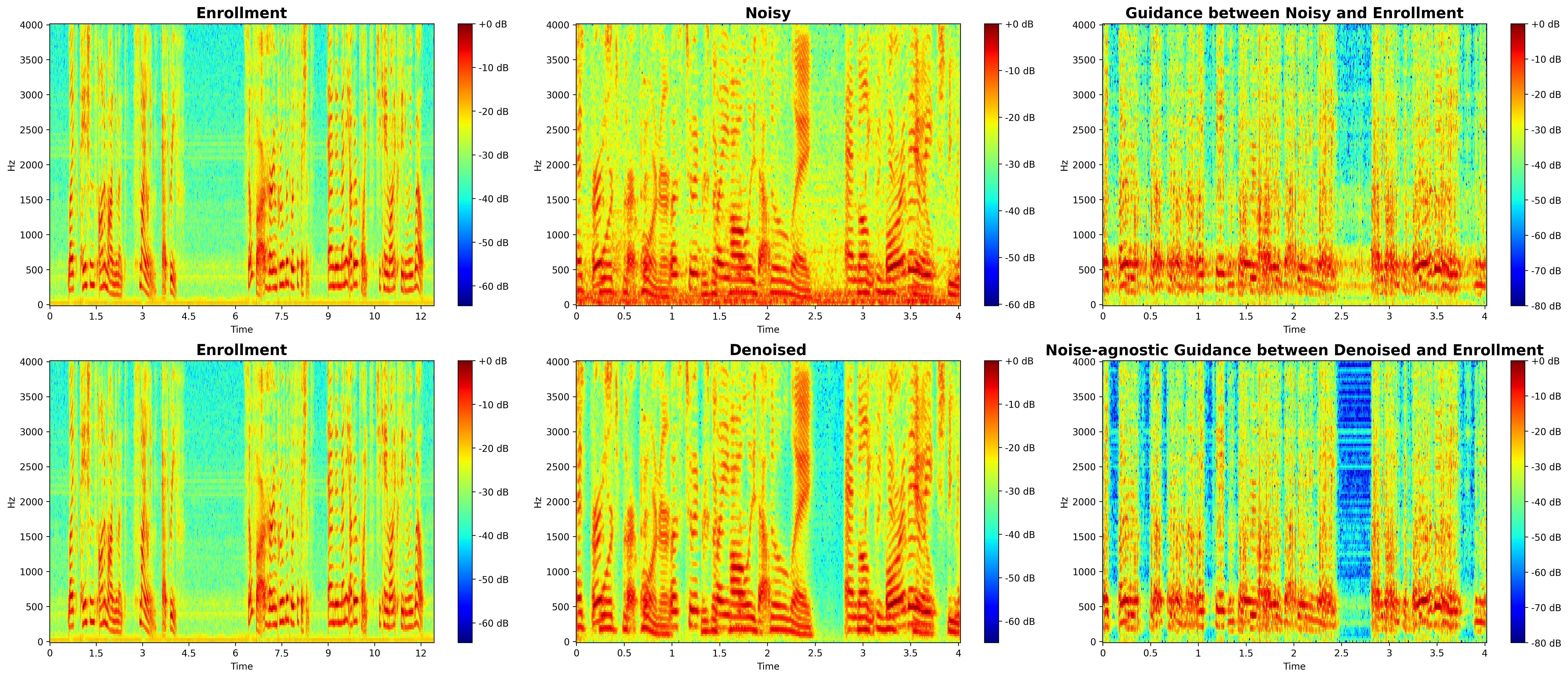}
\vspace{-0.3cm}
\caption{Noise-agnostic enrollment guidance analysis. The top row shows the enrollment guidance from direct context interaction between enrollment and noisy speech, while the bottom row shows the resulting spectrogram with the proposed noise-agonistic enrollment guidance.}
\label{fig:fig2}
\end{figure*}

\vspace{-0.2cm}  
\section{Experiments and Results}
\label{sec:res}
\subsection{Datasets}

All our experiments are performed on the Libri2Mix\cite{libri2mix} dataset, specifically using the `mix\_both' condition, 
which contains mixtures of a target speaker, one interfering speaker, and background noise. For clarity, 
we refer to this condition as `2-speaker + noise'. The training set includes 13,900 utterances from 251 speakers,
while both the development and test sets contain 3,000 utterances from 40 speakers each, with all mixtures 
simulated in the `minimum' mode. Note that only the first speaker is taken as the target speaker during 
all training mixture data simulation, and all mixtures are resampled to 8~kHz, unless otherwise specified.

\vspace{-0.3cm}  
\subsection{Models}

\textbf{GTCRN}\cite{rong2024gtcrn} consists of an encoder, a grouped dual-path recurrent neural network (G-DPRNN) module, 
and a decoder. The input mixture is first transformed from the STFT domain to ERB bands before 
being fed into the encoder. The encoder is composed of two convolutional blocks followed by 
three GT-Conv blocks. The decoder has a symmetric structure to the encoder, where convolutional 
blocks are replaced with transposed convolutional blocks. Finally, the output is converted back 
from the ERB domain to the STFT domain.

\textbf{SEF-PNet}\cite{sefpnet}, used as the competitive embedding-free TSE baseline, consists of an encoder, a decoder, 
a Temporal Convolutional Network (TCN) module, a PyramidBlock, and a Deconv2d layer. 
It is worth noting that the SEF-PNet baseline used in this study is a simplified version, where the iterative feature 
integration (IFI) block from the original design \cite{sefpnet} is removed to ensure a fair comparison with the proposed 
LGTSE and D-LGTSE. All other details remain identical to those in \cite{sefpnet}.

\textbf{CIE-mDPTNet} \cite{yangxue}, which has achieved state-of-the-art performance on TSE tasks, 
is also included as a baseline to further validate the effectiveness and generalization 
of our proposed methods. The detailed architecture can be found in \cite{yangxue}.


\vspace{-0.3cm}  
\subsection{Configurations}
We use a Hanning analysis window for STFT with a window length of 32 ms and a shift of 8 ms. The model is trained using the Adam optimizer with an initial learning rate of 0.0005. The learning rate is adjusted by multiplying it by 0.98 every two epochs for the first 100 epochs and by 0.9 for the last 20 epochs. Gradient clipping is applied to limit the maximum L2-norm to 1. The training procedure lasts up to 150 epochs.
For evaluation, we report SISDR (dB)\cite{sisdr}, PESQ\cite{pesq}, and STOI (\%)\cite{stoi}.

\vspace{-0.2cm}  
\subsection{Results}
\subsubsection{Overall Results}

\begin{table}[!t]
\centering
\vspace{-0.3cm}  
\caption{Overall Results. LGTSE and D-LGTSE are both with the SEF-PNet backbone network, while 
D-LGTSE-mDPTNet denotes the D-LGTSE framework equipped with the CIE-mDPTNet backbone.\\}
\label{tab:table1}
\begin{tabular}{@{}l@{\hskip 1em}lccc@{}}
\toprule
\textbf{ID} & \textbf{Methods} & \textbf{SI-SDR} & \textbf{PESQ} & \textbf{STOI} \\
\midrule
E0 & Unprocessed & -2.03 & 1.43 & 64.65 \\
E1 & SEF-PNet \cite{sefpnet} & 7.43 & 2.14 & 80.31 \\
\midrule
E2 & LGTSE & 7.88 & 2.21 & 81.27 \\
E3 & D-LGTSE (Concat) & 7.96 & 2.24 & 81.37 \\
E4 & D-LGTSE (On-the-fly) & 8.10 & 2.28 & 81.80 \\
E5 & D-LGTSE (Offline) & \textbf{8.32} & \textbf{2.30} & \textbf{82.28} \\
\midrule
F0 & CIE-mDPTNet \cite{yangxue} & 10.87 & 2.73 & 87.26 \\
F1 & D-LGTSE-mDPTNet (Offline)  & \textbf{11.70} & \textbf{2.86} & \textbf{88.83} \\
\bottomrule
\end{tabular}
\vspace{-0.4cm}  
\end{table}

Table~\ref{tab:table1} presents the overall performance of LGTSE and D-LGTSE compared 
with the two strong baselines. From E1 to E5, it can be observed that both LGTSE and D-LGTSE achieve 
consistent improvements over the baseline SEF-PNet across all evaluation metrics. 
Specifically, LGTSE outperforms SEF-PNet by 0.45 dB in SI-SDR and improves PESQ and STOI from 2.14/80.31 to 
2.21/81.27, demonstrating the effectiveness of noise-agnostic enrollment guidance. 
Among the three distortion-aware data usage mechanisms, D-LGTSE (Offline) yields the best overall performance, 
boosting the metrics to 8.32/2.30/82.28. This can be attributed to the fact that, during joint training after 
unfreezing, the denoiser in the concatenation and on-the-fly mechanisms becomes increasingly effective, which 
reduces the degree of residual distortion and limits the model’s exposure to challenging conditions. In 
contrast, the offline strategy stores distorted speech in advance, ensuring that such data remain available 
during the whole training and thus preserving the robustness benefits of distortion-aware learning. 

Beyond SEF-PNet, we further validate our framework on the stronger CIE-mDPTNet backbone. 
Although our implemented CIE-mDPTNet (F0) baseline already achieves SOTA performance, 
equipping it with our method, the proposed D-LGTSE-mDPTNet (Offline) delivers additional, 
sizable gains to 0.83/0.13/1.57 in SI-SDR/PESQ /STOI. Note that absolute metrics under the CIE-mDPTNet 
backbone exceed those with SEF-PNet mainly because CIE-mDPTNet incurs about 3× higher computational cost 
than our CNN-based SEF-PNet backbone (as summarized in Table~\ref{tab:model_complexity}). 
These results show that D-LGTSE remains highly effective when integrated with stronger backbones, 
demonstrating the generalizability of noise-agnostic enrollment guidance and 
distortion-aware training across architectures.

\vspace{-0.3cm}  
\begin{table}[!t]
\centering
\caption{Model size and computational complexity.}
\label{tab:model_complexity}
\begin{tabular}{lcc}
\toprule
\textbf{Model} & \textbf{Params (M)} & \textbf{MACs (G/s)} \\
\midrule
GTCRN        & 0.05  & 0.03 \\
\midrule
SEF-PNet     & 6.08  & 8.50 \\
D-LGTSE     & 6.13  & 8.53 \\
\midrule
CIE-mDPTNet  & 2.87  & 22.25 \\
D-LGTSE-mDPTNet  & 2.92  & 22.28 \\
\bottomrule
\end{tabular}
\vspace{-0.3cm}  
\end{table}

\subsubsection{Ablation Study of Model Training Strategy}

Table~\ref{tab:table3} reports the ablation study on different training strategies 
for D-LGTSE (Offline). In S0, the GTCRN and backbone are pretrained separately, 
and their combination yields limited performance, indicating that simple stacking 
without joint optimization is suboptimal.
In S1, the pretrained GTCRN is integrated as the front-end, while the backbone 
is trained from scratch. This closer coupling between modules improves the results 
to 8.02/2.26/81.41, showing the benefit of tighter integration. Finally, in E5, 
both GTCRN and the backbone are jointly optimized through a two-stage training scheme, 
which further enhances performance and achieves the best results.
These results demonstrate that progressively deeper integration and end-to-end 
joint training are crucial for fully exploiting the potential of distortion-aware 
learning in D-LGTSE.

\begin{table}[!t]
\centering
\vspace{-0.3cm}  
\caption{Different model training strategy of D-LGTSE(Offline).}
\label{tab:table3}
\begin{tabular}{@{}l@{\hskip 1em}lccc@{}}
\toprule
\textbf{ID} & \textbf{Training Method} & \textbf{SI-SDR} & \textbf{PESQ} & \textbf{STOI} \\
\midrule
S0 & GTCRN* + backbone* & 7.60 & 2.15 & 80.64 \\
S1 & GTCRN* + backbone & 8.02 & 2.26 & 81.41 \\
E5 & Two-stage Training   & \textbf{8.32} & \textbf{2.30} & \textbf{82.28} \\
\bottomrule
\end{tabular}
\vspace{-0.2cm}  
\end{table}

\subsubsection{Visualization of Noise-agnostic Enrollment Guidance}

Fig.\ref{fig:fig2} presents the effect of noise-agnostic enrollment guidance. 
The top row displays the enrollment spectrogram alongside noisy speech and 
their guidance obtained through direct context interaction. The bottom row shows the enrollment 
with denoised speech and the corresponding noise-agnostic guidance. 
By comparing the enrollment under noisy and denoised conditions, the denoising capability of 
GTCRN is clearly observed. Moreover, a comparison of the guidance spectrogram reveals that 
leveraging denoised speech for context interaction effectively suppresses noise components, 
leading to cleaner and more reliable guidance.

\section{Conclusion}
\label{sec:conclusion}

In this work, we proposed LGTSE and its distortion-aware extension D-LGTSE for target speech 
extraction in noisy multi-speaker scenarios. LGTSE leverages noise-agnostic enrollment guidance 
to prevent corruption of enrollment information, while D-LGTSE enhances robustness against 
speech distortion through distortion-aware training. A two-stage training strategy is 
further introduced to jointly optimize denoising and extraction. Experiments on the 
Libri2Mix 2-speaker+noise benchmark confirm consistent and significant improvements 
over strong baselines, validating the effectiveness of our approach. In future work, 
we will extend the proposed framework to broader TSE applications and larger, 
more diverse datasets to further verify its generalization.




\bibliographystyle{IEEEbib}
\bibliography{refs}
\end{document}